\documentclass[pdflatex,sn-mathphys-num]{sn-jnl}

\usepackage{graphicx}%
\usepackage{multirow}%
\usepackage{amsmath,amssymb,amsfonts}%
\usepackage{amsthm}%
\usepackage{mathrsfs}%
\usepackage[title]{appendix}%
\usepackage{xcolor}%
\usepackage{textcomp}%
\usepackage{manyfoot}%
\usepackage{booktabs}%
\usepackage{algorithm}%
\usepackage{algorithmicx}%
\usepackage{algpseudocode}%
\usepackage{listings}%

\theoremstyle{thmstyleone}%
\theoremstyle{thmstyletwo}%

\theoremstyle{thmstylethree}%

\begin{document}

\title[Article Title]{The RIGID Framework: Research-Integrated, Generative AI-Mediated Instructional Design}


\author[]{\fnm{Yerin} \sur{Kwak}}\email{kwak@berkeley.edu}

\author[]{\fnm{Zachary A.} \sur{Pardos}}\email{pardos@berkeley.edu}

\affil[]{\orgdiv{Berkeley School of Education}, \orgname{University of California, Berkeley}, \orgaddress{\postcode{94720}, \state{CA}, \country{USA}}}


\abstract{Instructional Design (ID) often faces challenges in incorporating research-based knowledge and pedagogical best practices. Although educational researchers and government agencies emphasize grounding ID in evidence, integrating research findings into everyday design workflows is often complex, as it requires considering multiple context-specific demands and constraints. To address this persistent gap, this paper explores how research in the learning sciences (LS) can be systematically integrated across ID workflows and how recent advances in generative AI can help operationalize this integration. While ID and LS share a commitment to improving learning experiences through design-oriented approaches in authentic contexts, structured integration between the two fields remains limited, leaving their complementary insights underutilized. We present RIGID (Research-Integrated, Generative AI-Mediated Instructional Design), a unified framework that integrates LS research across ID workflows spanning analysis, design, implementation, and evaluation phases, while leveraging generative AI to mediate this integration at each stage. The RIGID framework provides a systematic approach for enabling research-integrated instructional design that is both operational and context-sensitive, while preserving the central role of human expertise.}


\maketitle

\section{Introduction}\label{sec-intro}

As technology becomes increasingly integrated into education, policymakers and researchers have emphasized the importance of grounding instructional design (ID) in research evidence and established best practices. The U.S. Department of Education has issued guidelines for the development of AI-enabled educational products, highlighting the need to anchor design decisions—ranging from individual features to overarching frameworks—in well-established scientific rationales to enhance student learning \cite{cardona2023artificial,cardona2024designing}. Moreover, studies show that incorporating research findings into the design of learning experiences enhances instructional quality and student outcomes \cite{lierman2019developing,abuhassna2024exploring,caskurlu2025data,liu2025effectiveness}. 

Despite broad agreement on the value of evidence-informed design, integrating research-based insights into ID remains challenging. Existing ID approaches offer limited mechanisms to incorporate research findings into design practice and play a limited role in theory building \cite{yanchar2009beyond}. The learning sciences (LS) community has also long recognized that LS research has not readily translated into actionable guidance for everyday instructional design work~\cite{nathan2010learning,fishman2004creating}. One explanation is that research and practice operate with different interests and knowledge needs \cite{nathan2010learning,mcintyre2005bridging}. Educational research often produces propositional knowledge, or generalized insights intended to apply across contexts, whereas instructional decisions must accommodate multiple contextual factors, including classroom dynamics and institutional environments  \cite{koedinger2013instructional}. As a result, even well-established best practices require adaptation to local conditions and needs. In practice, instructional decisions often draw on instructors’ experiential and professional knowledge \cite{caskurlu2025data,yanchar2010struggling,rowland1993designing}, which may not always be systematically connected to broader research findings \cite{khalil2016applying,levinson2010where}.

To bridge this divide, prior work has proposed institutional mechanisms to connect research and practice, including collaborative school–researcher partnerships, professional development networks, structured knowledge-exchange programs \cite{mcintyre2005bridging,schlicht2024bridging}, and research–practice partnerships (RPPs) \cite{coburn2013practice,coburn2016research}. Although these initiatives underscore the importance of sustained engagement, they often rely on resource-intensive infrastructures and voluntary participation from instructors who already face limited time and competing responsibilities, limiting their scalability and long-term sustainability \cite{foster2014barriers}.

More recently, efforts have sought to embed research insights directly into ID activities. Some approaches leverage empirical analyses of learner data and interactions and research-informed principles to support design generation and iterative refinement, thereby improving student learning outcomes \cite{mah2025co,yan2021including,liu2025effectiveness}. Beyond design processes themselves, other scholars have emphasized the importance of implementing and sustaining research-based innovations within broader organizational and socio-technical systems, highlighting the ongoing work required to align innovations with contextual factors \cite{chen2024framework}. 

Together, these strands indicate that research–practice integration is not limited to initial design decisions but spans multiple phases of ID—from analysis and design to implementation and evaluation. Yet achieving such lifecycle-wide integration poses a deeper challenge. Instructional design decisions must simultaneously consider multiple contextual factors, making their coordination a combinatorial process that imposes substantial cognitive and logistical demands on instructors \cite{koedinger2013instructional}. This highlights the need for a framework that can systematically operationalize such integration throughout the full ID lifecycle without imposing substantial additional burdens on instructors.

In response, we propose RIGID (Research-Integrated, Generative AI-Mediated Instructional Design), a framework that integrates research in the learning sciences into ID processes by leveraging generative AI as a mediating tool. While ID and LS share a commitment to improving learning through design-oriented approaches in authentic contexts and emphasize the importance of connecting theory and practice \cite{merrill1996reclaiming,anderson2012design}, the two fields have largely evolved along separate trajectories, with limited integration across their theoretical and methodological traditions. Historically, ID has emphasized structured processes and practical methods for developing and implementing instruction \cite{saccak2022down}, whereas LS has focused on designing context-sensitive interventions and developing empirically grounded, theory-driven accounts of how people learn in complex, real-world settings \cite{hoadley2018short}. Integrating their complementary strengths provides a more robust foundation for research-integrated ID. However, achieving such integration in practice requires coordinating multiple instructional and contextual considerations, leaving few viable mechanisms for translating research insights into ID practices. Recent advances in generative AI create new opportunities to operationalize such integration while easing the burdens placed on instructors and researchers. By synthesizing the strengths of both traditions and leveraging AI as a mediating tool, RIGID offers a comprehensive and operational framework for connecting research and practice.

\section{Instructional Design and Learning Sciences}
\label{id-ls}
ID and LS share broader commitments and aims \cite{merrill1996reclaiming,anderson2012design}, yet the two fields have evolved along largely separate trajectories. We begin by examining how they differ in their core focus, methodological approaches, and the ways their work is applied in real-world educational contexts. We then discuss the limitations of each field's approach to bridging research and practice and show how their complementary strengths can help address these challenges.

\subsection{Core Focus}
Early ID scholars conceptualized ID as a systematic process for translating principles of instruction and learning into instructional materials and activities \cite{merrill1996reclaiming,smith2004instructional}. In this tradition, instructional designers apply methodologies grounded in instructional theories and models to design and develop content and solutions that facilitate the acquisition of new knowledge and skills \cite{saccak2022down}. Epistemologically, many early ID approaches assumed that knowledge can be transmitted through well-structured instruction. Within this view, the instructor’s role was to possess, organize, and convey knowledge, whereas learners are expected to grasp, replicate, and build on the information they receive \cite{heaster2020popular}. This stance contributed to the prescriptive nature of ID, which emphasized clearly defined objectives, observable outcomes, sequenced procedures, and systematic alignment of instruction with learner needs and contexts \cite{gagne1985conditions,dick2005systematic}.

On the other hand, LS focuses on understanding how people learn and how to support learning in authentic contexts \cite{hoadley2018short}. Grounded in developmental and constructivist theories of learning \cite{nathan2010learning}, LS positions the learner at the center of the learning process, emphasizing how their experiences and interactions shape the knowledge they construct. From this epistemological perspective, learners actively construct meaning through engagement with their environment, tools, and social interactions, and the instructor’s role is not to transmit information but to design environments, tasks, and questions that guide and scaffold learners’ meaning-making processes \cite{piaget1970,vygotsky1978}. Consistent with this perspective, LS researchers do not rely primarily on controlled laboratory studies; instead, they investigate learning in real-world settings to understand how context shapes learning and to design innovative solutions in collaboration with educators, learners, and other stakeholders \cite{coburn2016research,ishimaru2022transforming}. As a result, LS has been more research-oriented and primarily descriptive, with the goal of developing and refining theories that explain learning in context. In this tradition, design-based research (DBR) has emerged as a key methodological approach for investigating the mechanisms of learning and the influence of context \cite{brown1992design}.

\subsection{Methodological Approaches}
Early ID models posited that following systematic design principles would reliably produce effective instruction \cite{saccak2022down}. As a result, ID models have traditionally emphasized procedural and practical guidance for designing and implementing instruction and related resources. Among existing models, the Analyze, Design, Develop, Implement, and Evaluate (ADDIE) model remains foundational in the field \cite{branch2009instructional}. However, this linear, systematic approach has been criticized for oversimplifying the complexity and dynamism of contemporary learning environments, prompting the emergence of alternative models such as the Successive Approximation Model \cite{allen2012leaving} and Dick \& Carey model \cite{dick2005systematic}, which incorporate more flexible and iterative design perspectives. Despite their differences, most ID models share a common structure that includes analysis, design and development, implementation, and evaluation, underscoring ID’s focus on guiding the design of instruction through clearly articulated stages.

Among the various methodologies in learning sciences \cite{sawyer2022introduction,yoon2017learning}, Design-Based Research (DBR) has been widely used. We use the term DBR to encompass related approaches that have been labeled variously in the literature \cite{wang2005design}, including design experiments \cite{brown1992design}, design research \cite{edelson2002design}, and design-based research \cite{anderson2012design}. Despite the different terminology, these approaches share the same underlying goals and methods: DBR is a systematic methodology for improving educational practices and addressing real-world problems through iterative cycles of analysis, design, implementation, and evaluation \cite{anderson2012design}. Its dual purpose is to generate theoretically grounded solutions to practical problems and to refine or develop theories and design principles based on insights that emerge from authentic settings \cite{design2003design,sandoval2004design}. In each cycle, researchers identify a problem and analyze the contextual conditions that shape it, design interventions informed by theory and prior research, implement these interventions where the problem is situated, and evaluate their effectiveness \cite{Scott2020DesignBasedResearch}. Each cycle concludes with reflection that leads to further refinement of both the intervention and the theoretical understanding. The outcomes of DBR therefore include not only practical solutions but also theoretical contributions that describe mechanisms of learning in context.

\subsection{Applications in Educational Contexts}

ID is applied across sectors-—including industry, the military, and education-—but within educational settings it operates primarily at the classroom and institutional levels. Because ID emphasizes measurable learning outcomes, a core concern is ensuring that instructional materials work in practice. This requires instructors and designers to make a wide range of enactment decisions, such as determining the sequence and duration of activities, selecting materials, adjusting task difficulty, structuring interaction patterns, and planning assessment procedures \cite{lee2025collaborative}. To support these decisions, instructors often collaborate with instructional designers to plan, refine, and evaluate instruction \cite{iqbal2025codesigning,gronseth2024bridging}. Instructional designers typically apply design principles and models to educational projects \cite{ritzhaupt2015knowledge,kenny2005review}, assist faculty in improving their teaching practices \cite{miller2016finding,pan2009exploring}, and provide both technological and pedagogical support \cite{park2017refining,you2010instructional}. Collaborative design approaches—-in which designers and instructors jointly plan and iteratively refine course materials-—have been shown to enhance faculty capacity to integrate technology into instruction \cite{scoppio2017mind} and to support the development of high-quality online and blended courses \cite{olesova2019impact}.

In learning sciences, DBR has typically been applied in classroom and institutional settings to design educational interventions while generating theoretical insights. For example, DBR research on students’ understanding of statistical distributions identified recurring learning patterns, such as overreliance on small samples and confusion between spread and density, and demonstrated that a growing-samples activity effectively supported students in recognizing stable distributional features \cite{bakker2004design}. The study produced a domain-specific instructional theory grounded in diagrammatic reasoning and principles of Realistic Mathematics Education. 

DBR has also been used beyond individual classrooms. At the institutional level, the Center for Innovation in Learning and Student Success (CILSS) at the University of Maryland applied DBR to integrate adaptive learning into the curriculum for campus-wide initiative to improve course success rates \cite{ford2017using}. Through their research, they identified systemic barriers, refined curricular structures, and integrated new technologies that supported student success. Similarly, DBR projects in Jordan, Lebanon, and Saudi Arabia illustrate how school-level improvement efforts can surface actionable about teacher capacity, professional development needs, and contextualized approaches to fostering critical competencies \cite{akkary2019rethinking}. 

Although less common, LS design approaches have also been extended to the policy level. By conceptualizing education policy as a designed artifact, researchers have shown that policy design can involve iterative cycles of theorizing, observing implementation, analyzing discrepancies, and proposing revisions \cite{cobb2012analyzing}. This approach helps anticipate how policies will be interpreted and enacted in local contexts, understand why they succeed or fail, identify misalignments between policy intentions and classroom realities, and incorporate empirically grounded recommendations into instructional design and implementation.

\subsection{Limitations of ID and LS and the Need for Integration}

Despite their contributions, neither ID nor LS provides, on its own, a fully developed mechanism for reliably connecting research to practice. ID has long been criticized for offering limited mechanisms to integrate contemporary research findings into design practice and for lacking a robust methodological role in theory building \cite{yanchar2009beyond}. As a result, the incorporation of research-based principles frequently depends on individual designers’ interpretations, who must translate theoretical propositions into concrete instructional decisions such as task structure, sequencing, materials, and assessment. Even when they do so successfully, their work is further constrained by the need for sustained professional development \cite{muljana2021utilizing}.

Conversely, research in LS often does not readily translate its innovations and theories into actionable guidance for classroom implementation \cite{nathan2010learning}. While LS excels at explaining how learning occurs in authentic contexts, it offers limited support for turning insights derived from its research into implementable, context-sensitive instructional procedures. LS research findings do not typically prescribe the specific, local design decisions that instructors must make—such as calibrating task difficulty, structuring interactions, or designing assessment routines. These design decisions are essential for ensuring that research-based innovations can be enacted and sustained. DBR was originally proposed as a way to address some of these challenges by integrating iterative design with theory building. However, it remains primarily a research methodology and is rarely used in everyday ID practice. This is likely because DBR emphasizes iterative design, in-situ testing, and collaboration among multiple participants, which requires considerable time, resources, and financial investment \cite{anderson2012design}. It also mainly focuses on an innovation’s use within the study context and not necessarily on broader external factors, such as school systems and policy contexts, that enable its sustainable use \cite{fishman2004creating}. Thus, many LS-driven interventions remain tightly bound to the research projects in which they were developed and fail to persist once external support ends \cite{chen2024framework}.

These limitations illustrate that neither ID nor LS alone provides an adequate mechanism for bridging research and practice. Yet they also reveal the complementary strengths of the two fields: ID provides the systematic workflows and procedural insights, while LS offers approaches for designing context-sensitive interventions and provides theoretical and contextual explanations of how learning unfolds through these interventions in authentic settings. When considered together, these strengths show the potential for an integrated approach—one capable of supporting research-integrated, contextually grounded ID.

\subsection{From Previous Efforts to Emerging Opportunities in Generative AI}

Recent efforts have sought to bridge research and practice by embedding research insights and empirical analyses directly into design and development processes, with the aim of developing more rigorous and research-aligned learning materials. For example, empirical analyses of learner interaction and performance data have informed design decisions and supported iterative refinement \cite{mah2025co,yan2021including,liu2025effectiveness}. These approaches have advanced the integration of research and ID practices, particularly during design and development, but they offer limited guidance for how these interventions should be implemented, evaluated, and adapted across different contexts \cite{liu2025effectiveness}. At the same time, other scholars in the learning sciences have noted that research-informed innovations do not operate in isolation but must be embedded within organizational and socio-technical systems, where their success depends on ongoing adaptation through collaboration among teachers, students, researchers, and administrators \cite{chen2024framework}. Together, these strands suggest that successful research–practice integration cannot remain confined to isolated stages of ID; rather, it must extend across the full lifecycle—from problem analysis and design prototyping to implementation and evaluation. 

However, achieving such lifecycle-wide integration of ID workflows and LS research presents a deeper structural challenge. Translating between these two levels is not a linear mapping but a combinatorial process: it requires simultaneously coordinating theoretical principles, learner characteristics, institutional constraints, policy conditions, and available resources. As these variables interact, the space of possible design decisions expands rapidly \cite{koedinger2013instructional}. Sustaining such integration across iterative cycles of analysis, design, implementation, and evaluation imposes substantial cognitive demands on human designers and instructors. Addressing this problem requires a mediating mechanism capable of synthesizing heterogeneous knowledge sources, managing complex constraints, and supporting iterative refinement at scale.

Recent advances in generative AI create new possibilities for such mediation that were previously difficult to achieve. Unlike static guidelines, manual coordination processes, or rule-based design supports, generative AI can synthesize heterogeneous inputs and generate context-sensitive outputs. In doing so, it helps manage the combinatorial complexity that often places substantial demands on individual instructors. In this sense, AI functions as a mediating mechanism that can offload the coordination of complex, multi-level knowledge while preserving the instructor's central role in interpretation, judgment, and contextual adaptation. Building on these perspectives, we propose the RIGID framework as a unified approach for research-integrated, generative AI–mediated instructional design.

\section{The RIGID Framework}

\begin{figure}[t]
\centering
\includegraphics[width=1\textwidth]{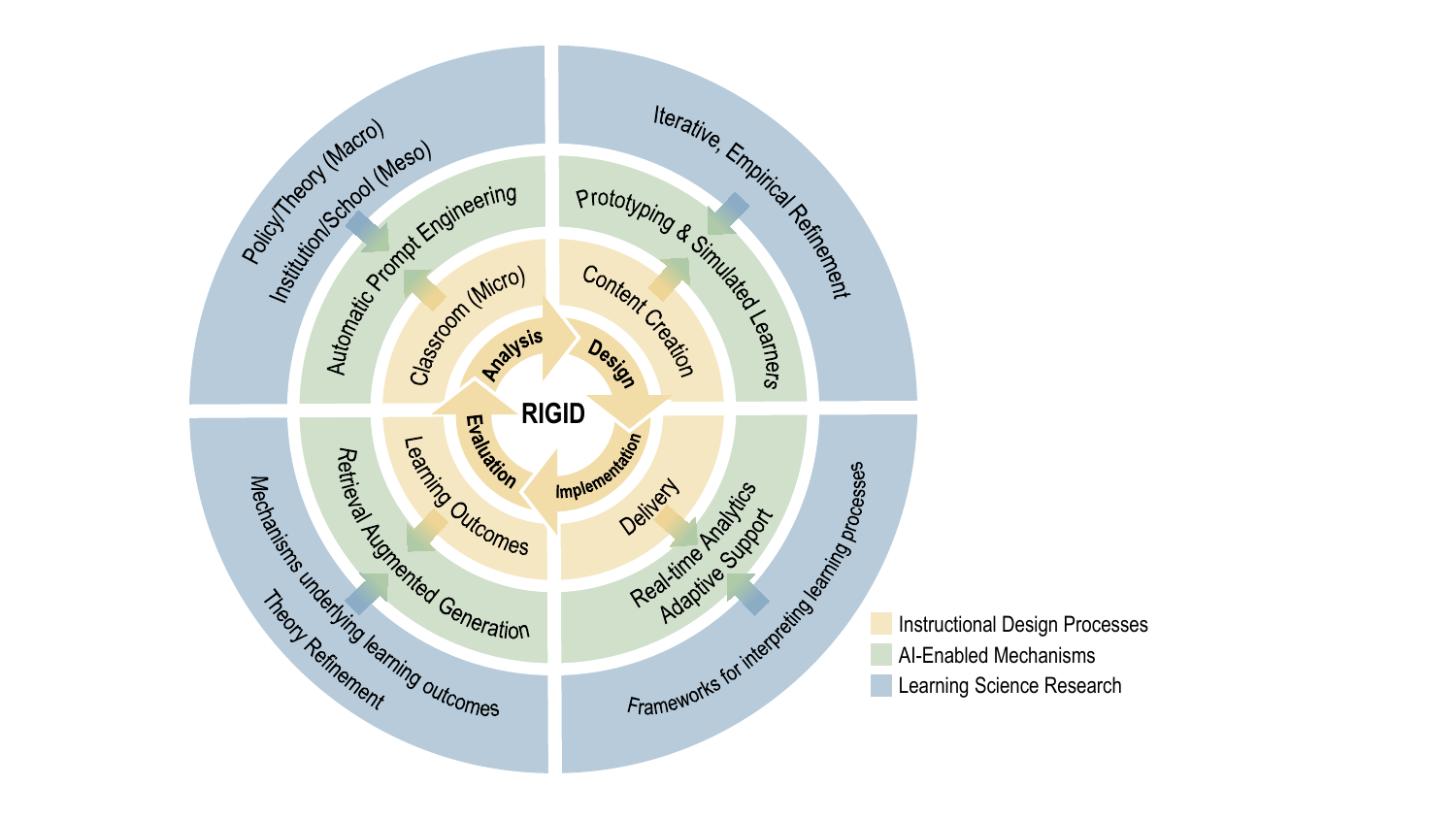}
\caption{\textbf{Framework for research-integrated instructional design.} The RIGID framework integrates Learning Sciences perspectives into instructional design processes across four iterative phases, with AI serving as a mediating mechanism. The inner ring represents core instructional design activities, the middle ring illustrates how AI supports their integration, and the outer ring depicts contributions from the Learning Sciences. Together, the framework shows an iterative cycle in which insights from each phase inform and refine subsequent instructional design decisions.}
\label{fig1}
\end{figure}

The RIGID framework is organized around a typical ID process—Analysis, Design, Implementation, and Evaluation. We describe how LS research perspectives can be incorporated into each phase of the ID process and how AI can support and operationalize the translation between LS insights and ID decisions. Consistent with how both ID and LS conceptualize design work, the framework reflects an iterative cycle in which insights generated during Evaluation inform subsequent rounds of Analysis and Design. Figure \ref{fig1} provides an overview of the framework.

\subsection{Analysis}

In ID, the analysis phase focuses on clarifying instructional goals, understanding learners’ characteristics, assessing available resources, and determining potential delivery formats \cite{branch2009instructional}. At this stage, instructors must contend with the inherent complexity of classroom life, which involves simultaneous and unpredictable demands that require immediate and context-dependent decisions \cite{doyle2013ecological}. Although this process provides a structured foundation for generating instruction, it typically focuses on the immediate instructional context where learners are situated. By itself, such analysis may not sufficiently account for broader institutional and systemic conditions or explicitly incorporate research-based insights that extend beyond the classroom.

RIGID expands the scope of analysis by incorporating Kozma’s (2003) framework of three contextual levels that shape pedagogical practice: micro, meso, and macro \cite{kozma2003technology}. At the micro level, classroom factors such as student characteristics, teacher expertise, learning objectives, and pedagogical constraints are central, aligning with the focus of traditional ID analysis. The meso level encompasses institutional conditions, including organizational goals, resources, and local policies that enable or constrain instructional activity. The macro level reflects the broadest context, including educational policies and theoretical perspectives that inform what counts as effective teaching and learning. Effective instructional practice emerges when all three factors interact in a mutually reinforcing way, rather than when one level unilaterally determines outcomes at another \cite{kozma2003technology}. 

LS research that examines the meso level provides instructors and instructional designers with an evidence-based understanding of specific organizational contexts, including which pedagogical approaches are effective, what kinds of support and resources are available, and which structural constraints must be considered at the outset of design \cite{ford2017using,akkary2019rethinking}. Rather than treating institutional factors as afterthoughts, meso-level insights enable designers to anticipate which forms of instructional innovation are viable and aligned with broader institutional goals. At this level, Centers for Teaching and Learning (CTLs), or similar institutional research units can serve as key contributors. Prior studies indicate that CTLs often function as institutional laboratories, identifying campus-specific research questions, analyzing data, designing and refining programs, and disseminating actionable knowledge about teaching and learning \cite{lieberman2005beyond}.

Similarly, incorporating macro-level policy research into the analysis phase offers clarity about which policies are effective under particular contextual conditions and ensures that instructional designs remain aligned with higher-level policy objectives \cite{cobb2012analyzing}. Considering macro-level influences early on also helps designers anticipate downstream consequences for learners and instructors, thereby supporting the development of instruction that is both contextually grounded and systemically coherent. In addition to policy insights, existing research and theories from LS and educational psychology can also inform the analysis phase by providing well-established principles of human learning \cite{bransford2005learning,harris2012apa,apa2015top20}. These bodies of work offer instructors and designers a research-grounded foundation for situating instructional decisions within broader theoretical frameworks.

Taken together, institutional research (meso-level), policy-oriented research, and principles from LS and educational psychology (macro-level) broaden the analysis phase beyond the immediate classroom (micro-level). Incorporating these layers early in the process enables designers to create instruction that is not only pedagogically sound but also contextually coherent and actionable, enabling the sustainable and scalable adoption of instructional designs~\cite{fishman2004creating}. It also helps prevent common mismatches—such as developing instructional innovations that are well aligned with instructors’ goals and students’ needs but are infeasible within institutional structures or not aligned with broader policy goals.

AI can support the analysis phase by transforming insights from the micro, meso, and macro levels—together with a system prompt that provides high-level, task-agnostic guidance for the model’s behavior and responses \cite{schulhoff2024prompt}—into structured prompts that inform subsequent design decisions. Recent advances in automatic prompt engineering (APE), techniques that automatically generate, evaluate, and select prompts, enable AI to produce multiple prompt variants and identify those that perform best for the intended purpose \cite{zhou2022large}. For example, when instructors provide micro-level information about their learners and classroom constraints, and LS research contributes meso- and macro-level insights, AI can integrate these inputs to generate and optimize prompts that can guide the creation of instructional materials aligned with all three contextual levels. This multi-level synthesis, combined with APE-driven optimization, sets the stage for AI to operationalize these insights in the design phase.

\subsection{Design}

The Design phase in our framework corresponds to what the ADDIE model separates into the Design and Development stages. In ID, the Design phase specifies the tasks, activities, assessments, and learning resources intended to support the learning objectives identified during Analysis \cite{branch2009instructional}. The Development phase typically follows as a distinct phase focused on producing and refining materials \cite{branch2009instructional}. In practice, however, these processes are tightly intertwined: material production often reshapes design decisions, and iterative refinement often alters initial plans. For this reason, RIGID integrates Design and Development into a unified phase.

During this phase, LS research findings define the boundaries of pedagogically and contextually reasonable options but do not prescribe deterministic instructional decisions or identify universally optimal strategies. Within these boundaries, decisions about which resources, environments, or task structures best support learning are inherently empirical. They must be iteratively explored and refined in relation to learner characteristics, institutional conditions, policy requirements, and available resources. Because these variables interact combinatorially, the space of possible design configurations expands rapidly \cite{koedinger2013instructional}. Exploring this high-dimensional space manually imposes substantial cognitive and logistical burdens on designers and instructors.

RIGID addresses this challenge by positioning generative AI as a coordination layer that operationalizes research-based constraints while enabling scalable exploration of the design space. By leveraging prompts and coordinating the micro-, meso-, and macro-level considerations identified during the Analysis phase, generative AI can generate multiple design alternatives within research-informed boundaries, enabling designers and instructors to systematically compare and evaluate alternative design configurations. Empirical studies suggest that when AI systems are provided with structured, context-rich inputs and carefully designed prompting strategies, their outputs more closely align with user intent and pedagogical objectives \cite{marvin2023prompt,kulkarni2024crafting}, and established instructional frameworks \cite{celik2026knowledge,celik2025co}. Generated materials may include lesson plans \cite{hu2025llm}, assessment items \cite{bhandari2024evaluating}, and adaptive tutoring content \cite{kwak2025adaptive}.

\begin{figure}[t]
\centering
\includegraphics[width=0.8\textwidth]{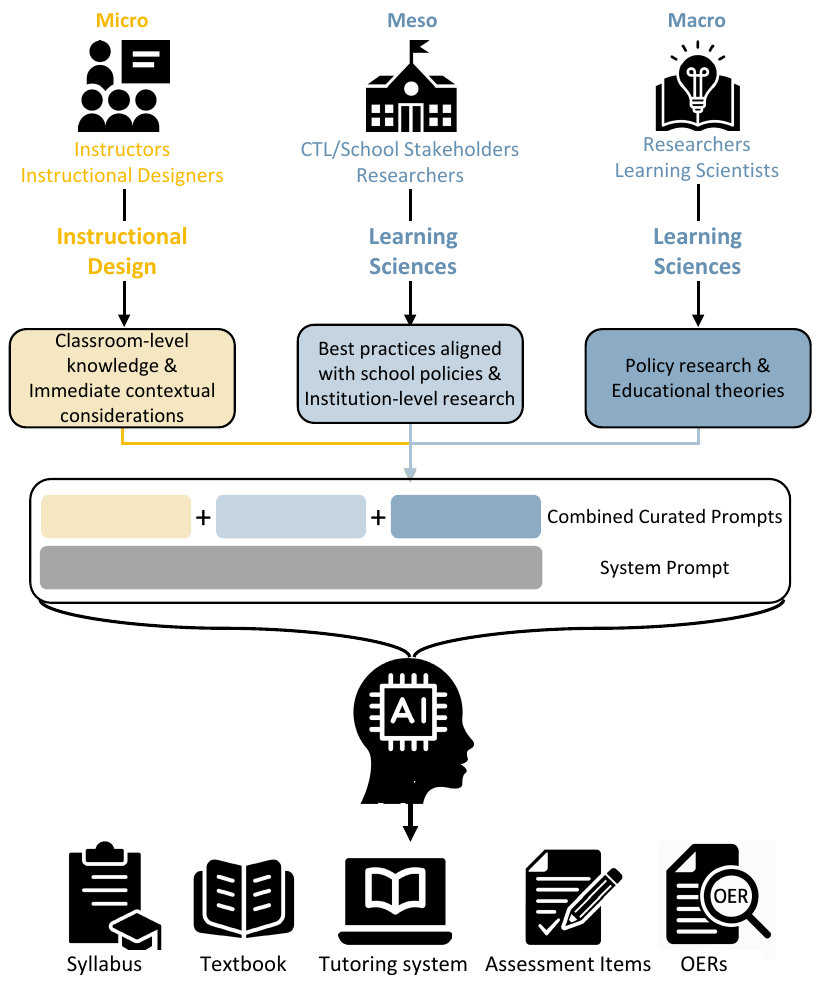}
\caption{\textbf{Operationalization of the analysis and design phases in the RIGID framework.} Micro-level insights from instructional design, together with meso- and macro-level insights from the learning sciences, are distilled into prompt components for the AI, which then optimizes prompts, prototypes materials, and simulates learners to create contextually grounded and pedagogically aligned instructional resources.}
\label{fig2}
\end{figure}

Beyond prototyping, generative AI can also support simulation-based evaluation prior to classroom deployment. Emerging work suggests that generative AI can simulate novice learners whose knowledge evolves over time \cite{jin2024teach}, generate responses aligned with specific knowledge states or cognitive traits \cite{lu2024generative,li2024evolving,jin2025teachtune}, and even emulate dynamic learner personas whose beliefs and abilities evolve over time \cite{nguyen2024simulating}. Within RIGID, such simulated learners can serve as exploratory tools for evaluating design prototypes: designers may use them to anticipate potential misunderstandings, examine how different learner profiles interact with materials, or estimate relative item difficulty across design variations. While simulated learners cannot substitute for validation with real students, they can narrow the design space prior to human testing, reducing ethical risks, temporal constraints, and resource demands associated with repeated classroom experimentation. These mechanisms transform combinatorial design complexity into a structured, research-aligned exploration process. Figure \ref{fig2} illustrates how the analysis and design phases of the RIGID framework can be operationalized.

Importantly, even with AI-assisted prototyping and simulation, instructors retain the central role in interpreting these outputs and making instructional decisions. AI-generated resources or simulated learner responses should not be regarded as prescriptive solutions; rather, they serve as draft materials that broaden the range of possibilities available to the instructor. They evaluate whether these drafts align with their learning goals, pedagogy, the specific needs of their students, and available resources. Based on this evaluation, they determine which elements to adopt, discard, or refine. AI can then generate revised versions, which instructors continue to adapt. This process ensures that AI augments—rather than replaces—teacher expertise, supporting instructional coherence while reducing the workload associated with iterative development.

\subsection{Implementation}

Within RIGID, implementation is not treated as a simple delivery stage but as a dynamic process in which micro-, meso-, and macro-level considerations continue to interact in real time. At the micro level, implementation resembles the focus of traditional ID: preparing the learning environment and engaging students with newly developed instructional materials \cite{branch2009instructional}. Instructors play a central role in facilitating instruction, setting the pace of learning, providing subject-matter expertise, monitoring students’ progress, and administering assessments. These activities form the practical foundation for helping students achieve the intended learning goals. 

Meso-level perspectives extend implementation beyond the classroom by situating instructional activity within institutional systems. At this level, implementation is examined in relation to institutional priorities and support structures—for example, whether instructional innovations align with institutional goals, how they are enacted and sustained across programs, whether resources are sufficient, and whether curricular adjustments are needed to maintain long-term viability. CTLs or institutional research units can use this information to provide targeted support, coordinate resources, and guide institutional-level improvements.

At the macro level, policy environments and theoretical frameworks from LS research provide interpretive lenses for understanding classroom events and interpreting student behaviors, misconceptions, and progress patterns. These broader perspectives help situate classroom practices within evolving research-based understandings of learning.

In practice, these levels are deeply interdependent yet rarely coordinated during implementation. Operationalizing such a multi-level perspective requires synthesizing data and performing real-time analytics to support instructional adjustment, which poses technical and organizational challenges \cite{gasevic2016we,wong2020review}. Learning platforms and institutional systems generate large volumes of data during instruction, but making sense of them in real time remains difficult. AI can support this multi-level perspective by synthesizing data and providing real-time analyses to different stakeholders, enabling instructors to make timely adjustments during instruction. Prior research has demonstrated the feasibility of AI-driven analytics for monitoring learner behaviors, surfacing patterns in learning processes, and generating instructional recommendations \cite{goslen2025llm,raj2024improved,li2025turning}. As a result, implementation becomes not merely a logistical task of delivery but a process of research-integrated practice in which instructional actions, institutional coordination, and research-informed interpretation are continuously aligned through data and analytics.

\subsection{Evaluation}
The evaluation phase extends beyond assessing whether instructional materials and processes were effective to examining how and why particular instructional outcomes occurred. In traditional ID, evaluation focuses on determining the effectiveness of instructional materials and processes by establishing evaluation criteria, selecting appropriate evaluation methods, and collecting evidence about learning outcomes \cite{branch2009instructional}. The primary goal is to determine whether the instruction achieved its intended objectives.

Integrating research from the learning sciences expands this evaluative focus by examining not only outcomes but also the mechanisms that produced those outcomes. Rather than solely asking whether an instructional approach worked, evaluation in RIGID investigates why it succeeded or failed by analyzing the instructional processes, learner interactions, and contextual conditions that shaped learning.

From a meso-level perspective, evaluation considers how instructional outcomes relate to institutional contexts, including whether outcomes align with institutional goals, whether available resources adequately support implementation, and how organizational conditions may have influenced learning processes and results. At the macro level, policy environments and theoretical frameworks from LS research provide explanatory frameworks for understanding the mechanisms underlying observed learning outcomes and instructional effects. 

AI can further support this evaluative process by synthesizing evaluation evidence across micro, meso, and macro contexts. Through such multi-level analysis, AI can surface patterns and relationships that are difficult to detect through manual analysis alone, enabling more systematic examination of the mechanisms underlying observed learning outcomes. Retrieval-Augmented Generation (RAG), which augments model outputs by retrieving relevant documents or knowledge \cite{lewis2020retrieval}, can further support this process by grounding AI-generated interpretations in relevant theoretical frameworks and contextual information. By connecting empirical evidence with research-informed knowledge, these analyses can produce explanations that are more aligned with existing theories and sensitive to contextual conditions. Thus, evaluation becomes not only a judgment of effectiveness but also a process of generating insights that can refine instructional practices and inform the development or advancement of learning theories. These insights can subsequently inform the next analysis phase, enabling iterative refinement of ID.

\section{Challenges and Considerations}

While the proposed framework offers a promising approach to integrating LS research into ID with the support of AI, several challenges should be addressed to help ensure successful adoption and sustainable implementation in real-world educational settings. 

\subsection{Hallucinations}
Despite rapid advancements, generative AI continues to exhibit hallucinations, confidently producing false or misleading information \cite{huang2025survey}, which remains a critical concern in educational contexts. Although recent work has shown that self-consistency, self-reflection, and similar methods can significantly reduce hallucination rates \cite{ji2023mitigating,pardos2024chatgpt}, error rates remain higher than desirable in many domains \cite{huang2025survey,li2024dawn}. These inaccuracies underscore the ongoing need for human oversight. Within the proposed framework, instructors and subject matter experts must review, revise, and validate AI-generated outputs to ensure factual accuracy, conceptual correctness, and pedagogical appropriateness before materials are implemented in the classroom. 

\subsection{The Role of Human Expertise}
Although generative AI excels at synthesizing information and generating outputs based on existing data, it does not yet demonstrate the capacity for original knowledge construction \cite{ding2025generative}. AI systems cannot replicate the epistemological processes involved in LS research or ID. As a result, human expertise remains essential for developing new insights, theory-informed interpretations, and contextually grounded instructional decisions—forms of professional judgment that cannot be produced merely by querying a model. Ensuring that human experts remain central to the framework is therefore critical for maintaining both rigor and relevance.

\subsection{Bias and Equity Considerations}
Despite recent improvements, AI continues to reflect biases present in their training data \cite{ferrara2023should}. Such biases can manifest in subtle but consequential ways, including generating examples that center dominant cultural norms or overlooking the needs of marginalized learners. In educational contexts, these patterns risk reinforcing inequities in instruction, assessment, and feedback. For instance, biased outputs may produce instructional materials that assume background knowledge not shared by all students, offer culturally narrow examples, or provide differential scaffolding quality based on linguistic or behavioral patterns in student inputs. To mitigate these risks, instructors must critically review AI-generated materials for cultural inclusiveness, accessibility, and fairness before implementing them in the classroom.

\subsection{Minimizing Instructor Friction and Supporting Adoption}
For the framework to be adopted at scale, the workflow for generating or adapting materials must be efficient, intuitive, and minimally disruptive to instructors’ existing practices. Even high-quality AI outputs may go unused if the process of eliciting them is perceived as time-consuming or cognitively burdensome. Minimizing instructor friction \cite{avgeriou2015reducing}—reducing barriers to use and enabling smooth integration—is therefore essential. Embedding AI-assisted design capabilities directly into platforms instructors already use, such as Learning Management Systems (LMSs), can streamline workflows and reduce the need for additional training. Early evidence supports the viability of such integrations: for instance, the generative AI authoring tool PromptHive, which supports instructors in generating and refining instructional materials, achieved a usability score of 89 out of 100 among subject matter experts \cite{reza2025prompthive}. Future system development should prioritize human-centered design, interoperability with existing tools, and low-effort interaction to ensure that instructors can leverage the framework without substantial additional workload.

\section{Towards Research-Integrated Instructional Design}
Bridging the longstanding divide between educational research and ID requires integrating LS research across the full ID lifecycle, as well as a mechanism capable of mediating between the two. 

The RIGID framework responds to this need by articulating how LS research can be systematically embedded within ID workflows, with generative AI serving as a mediating infrastructure across analysis, design, implementation, and evaluation. Importantly, this mediation does not diminish the role of human expertise. Instructors, designers, and researchers remain central to interpreting theory, exercising professional judgment, and ensuring contextual and ethical appropriateness.

Future research should examine how this framework functions across varied institutional settings and identify the conditions under which its limitations emerge. Empirical studies, critical analysis, and iterative refinement will be essential for evaluating its effectiveness and guiding its adaptation to diverse instructional challenges.

This work advances ID by introducing an operational, lifecycle-spanning approach that embeds theoretical and empirical insights into procedural workflows. At the same time, it contributes to the learning sciences by providing a structured pathway for translating research insights into sustained and contextually grounded educational practice.

\backmatter


\section*{Declarations}

\begin{itemize}
\item Funding: UCB Reimagining Higher Education Grant (ZP)
\item Conflict of interest: The authors declare that they have no competing interests.
\item Data availability: Not applicable
\item Materials availability: Not applicable
\item Code availability : Not applicable
\item Author contribution: ZP and YK contributed to the conceptualization, methodology, investigation, and visualization of the study. ZP acquired the funding. YK prepared the original draft, and ZP was responsible for reviewing and editing the manuscript.
\end{itemize}

\noindent

\bigskip




\bibliography{sn-bibliography}

@techreport{cardona2024designing,
  title     = {Designing for Education with Artificial Intelligence: An Essential Guide for Developers},
  author    = {Cardona, Miguel A. and Rodr{\'\i}guez, Roberto J.},
  year      = {2024},
  institution = {US Department of Education, Office of Educational Technology},
  type      = {Government Report}
}

@article{merrill1996reclaiming,
  title={Reclaiming instructional design},
  author={Merrill, M David and Drake, Leston and Lacy, Mark J and Pratt, Jean and ID₂ Research Group},
  journal={Educational Technology},
  pages={5--7},
  year={1996},
  publisher={JSTOR}
}

@article{lierman2019developing,
  title={Developing online instruction according to best practices},
  author={Lierman, Ashley and Santiago, Ariana},
  year={2019},
  journal ={Journal of Information Literacy}
}

@article{khalil2016applying,
  title={Applying learning theories and instructional design models for effective instruction},
  author={Khalil, Mohammed K and Elkhider, Ihsan A},
  journal={Advances in physiology education},
  year={2016},
  publisher={American Physiological Society Bethesda, MD}
}

@article{koedinger2013instructional,
  title={Instructional complexity and the science to constrain it},
  author={Koedinger, Kenneth R and Booth, Julie L and Klahr, David},
  journal={Science},
  volume={342},
  number={6161},
  pages={935--937},
  year={2013},
  publisher={American Association for the Advancement of Science}
}

@book{kozma2003technology,
  editor    = {Kozma, Robert B.},
  title     = {Technology, Innovation, and Educational Change: A Global Perspective},
  year      = {2003},
  publisher = {International Society for Technology in Education},
  address   = {Eugene, OR}
}

@inproceedings{reza2025prompthive,
  title={PromptHive: Bringing subject matter experts back to the forefront with collaborative prompt engineering for educational content creation},
  author={Reza, Mohi and Anastasopoulos, Ioannis and Bhandari, Shreya and Pardos, Zachary A},
  booktitle={Proceedings of the 2025 CHI Conference on Human Factors in Computing Systems},
  pages={1--22},
  year={2025}
}

@article{lieberman2005beyond,
  title={Beyond faculty development: How centers for teaching and learning can be laboratories for learning.},
  author={Lieberman, Devorah},
  journal={New directions for higher Education},
  volume={2005},
  number={131},
  year={2005}
}

@article{anderson2012design,
  title={Design-based research: A decade of progress in education research?},
  author={Anderson, Terry and Shattuck, Julie},
  journal={Educational researcher},
  volume={41},
  number={1},
  pages={16--25},
  year={2012},
  publisher={Sage Publications Sage CA: Los Angeles, CA}
}

@article{ford2017using,
  title={Using design-based research in higher education innovation.},
  author={Ford, Cristi and McNally, Darragh and Ford, Kate},
  journal={Online Learning},
  volume={21},
  number={3},
  pages={50--67},
  year={2017},
  publisher={ERIC}
}

@book{gagne1985conditions,
  author    = {Gagné, Robert M.},
  title     = {The Conditions of Learning and Theory of Instruction},
  edition   = {4th},
  year      = {1985},
  publisher = {Holt, Rinehart and Winston},
  address   = {New York}
}

@book{branch2009instructional,
  author    = {Robert Maribe Branch},
  title     = {Instructional Design: The ADDIE Approach},
  year      = {2009},
  publisher = {Springer},
  address   = {New York, NY},
  doi       = {10.1007/978-0-387-09506-6}
}

@book{allen2012leaving,
  title={Leaving ADDIE for SAM: An agile model for developing the best learning experiences},
  author={Allen, Michael and Sites, Richard},
  year={2012},
  publisher={Association for Talent Development}
}

@article{design2003design,
  title={Design-based research: An emerging paradigm for educational inquiry},
  author={Design-Based Research Collective},
  journal={Educational researcher},
  volume={32},
  number={1},
  pages={5--8},
  year={2003},
  publisher={Sage Publications Sage CA: Thousand Oaks, CA}
}

@article{sandoval2004design,
  title={Design-based research methods for studying learning in context: Introduction},
  author={Sandoval, William A and Bell, Philip},
  journal={Educational psychologist},
  volume={39},
  number={4},
  pages={199--201},
  year={2004},
  publisher={Taylor \& Francis}
}

@article{akkary2019rethinking,
  title={Rethinking design-based approaches for school-based improvement: The experience of the TAMAM Project},
  author={Akkary, Rima Karami and DeKnight, Jennifer},
  journal={International Journal of Educational Reform},
  volume={28},
  number={1},
  pages={99--121},
  year={2019},
  publisher={SAGE Publications Sage CA: Los Angeles, CA}
}

@techreport{cardona2023artificial,
  title={Artificial intelligence and the future of teaching and learning: Insights and recommendations},
  author={Cardona, Miguel A and Rodr{\'\i}guez, Roberto J and Ishmael, Kristina and others},
  year={2023},
  institution={US Department of Education, Office of Educational Technology},
  type={Government Report}
}

@inproceedings{marvin2023prompt,
  title={Prompt engineering in large language models},
  author={Marvin, Ggaliwango and Hellen, Nakayiza and Jjingo, Daudi and Nakatumba-Nabende, Joyce},
  booktitle={International conference on data intelligence and cognitive informatics},
  pages={387--402},
  year={2023},
  organization={Springer}
}

@article{kulkarni2024crafting,
  title={Crafting effective prompts: enhancing ai performance through structured input design},
  author={Kulkarni, Nilesh D and Tupsakhare, Preeti},
  journal={Journal of recent trends in computer science and engineering (JRTCSE)},
  volume={12},
  number={5},
  pages={1--10},
  year={2024}
}

@article{bransford2005learning,
  title={Learning theories and education: Toward a decade of synergy},
  author={Bransford, John and Vye, Nancy and Stevens, Reed and Kuhl, Pat and Schwartz, Daniel and Bell, Philip and Meltzoff, Andrew and Barron, Brigid and Pea, Roy D and Reeves, Byron and others},
  journal={Handbook of Educational Psychology (2nd Edition)},
  pages={95--pages},
  year={2005},
  publisher={Mahwah, NJ: Erlbaum.}
}

@inproceedings{kwak2025adaptive,
  author    = {Kwak, Yerin and Dunder, Nora and Viberg, Olle and Pardos, Zachary A.},
  title     = {Adaptive Tutoring Goes to Sweden: Machine Translation and Alignment of English OERs to a Swedish Calculus Course},
  booktitle = {Proceedings of the 12th ACM Learning @ Scale Conference},
  year      = {2025},
  note      = {Accepted for publication. To appear in L@S 2025, Palermo, Italy. ACM.}
}

@article{schulhoff2024prompt,
  author    = {Schulhoff, Sander and Ilie, Michael and Balepur, Nishant and Kahadze, Konstantine and Liu, Amanda and Si, Chenglei and Li, Yinheng and others},
  title     = {The Prompt Report: A Systematic Survey of Prompting Techniques},
  journal   = {arXiv preprint arXiv:2406.06608},
  year      = {2024},
  url       = {https://arxiv.org/abs/2406.06608}
}

@article{caskurlu2025data,
  author    = {Caskurlu, Secil and Yal\c{c}\i n, Yasin and Hur, Jaesung and Shi, Hui and Klein, James D.},
  title     = {Data-Driven Decision-Making in Instructional Design: Instructional Designers’ Practices and Strategies},
  journal   = {TechTrends},
  year      = {2025},
  pages     = {1--12},
  doi       = {10.1007/s11528-025-01077-x}
}

@article{yanchar2010struggling,
  author    = {Yanchar, Stephen C. and South, Joseph B. and Williams, David D. and Allen, Stephanie and Wilson, Brent G.},
  title     = {Struggling with theory? A qualitative investigation of conceptual tool use in instructional design},
  journal   = {Educational Technology Research and Development},
  volume    = {58},
  number    = {1},
  pages     = {39--60},
  year      = {2010},
  doi       = {10.1007/s11423-009-9129-6},
}

@article{rowland1993designing,
  author    = {Rowland, Gordon},
  title     = {Designing and Instructional Design},
  journal   = {Educational Technology Research and Development},
  volume    = {41},
  number    = {1},
  pages     = {79--91},
  year      = {1993},
  doi       = {10.1007/BF02297094}
}

@article{levinson2010where,
  author    = {Levinson, Anthony J.},
  title     = {Where is evidence-based instructional design in medical education curriculum development?},
  journal   = {Medical Education},
  volume    = {44},
  number    = {6},
  pages     = {536--537},
  year      = {2010},
  doi       = {10.1111/j.1365-2923.2010.03715.x},
}

@article{foster2014barriers,
  author    = {Foster, Robyn},
  title     = {Barriers and Enablers to Evidence‑Based Practices},
  journal   = {Kairaranga},
  volume    = {15},
  number    = {1},
  pages     = {50--58},
  year      = {2014},
}

@book{dick2005systematic,
  author    = {Dick, Walter and Carey, Lou and Carey, James O.},
  title     = {The Systematic Design of Instruction},
  edition   = {6th},
  year      = {2005},
  publisher = {Pearson/Allyn \& Bacon},
  address   = {Upper Saddle River, NJ},
  isbn      = {0205412742}
}

@book{harris2012apa,
  editor    = {Harris, Karen R. and Graham, Steve E. and Urdan, Tim E. and McCormick, Cynthia B. and Sinatra, Gale M. and Sweller, John E.},
  title     = {APA Educational Psychology Handbook (Volumes 1--3)},
  publisher = {American Psychological Association},
  address   = {Washington, DC},
  year      = {2012},
  isbn      = {978-1-4338-0210-9},
  note      = {3-volume set}
}

@article{ding2025generative,
  author    = {Ding, Amy Wenxuan and Li, Shibo},
  title     = {Generative AI lacks the human creativity to achieve scientific discovery from scratch},
  journal   = {Scientific Reports},
  volume    = {15},
  number    = {1},
  year      = {2025},
  doi       = {10.1038/s41598-025-93794-9},
  note      = {Published March 20, 2025}
}

@article{avgeriou2015reducing,
  author    = {Avgeriou, Paris and Kruchten, Philippe and Nord, Robert L. and Ozkaya, Ipek and Seaman, Carolyn},
  title     = {Reducing Friction in Software Development},
  journal   = {IEEE Software},
  volume    = {33},
  number    = {1},
  pages     = {66--73},
  year      = {2015},
  doi       = {10.1109/MS.2016.13},
}

@article{huang2025survey,
  author    = {Huang, Lei and Yu, Weijiang and Ma, Weitao and Zhong, Weihong and Feng, Zhangyin and Wang, Haotian and Chen, Qianglong and Peng, Weihua and Feng, Xiaocheng and Qin, Bing and Liu, Ting},
  title     = {A Survey on Hallucination in Large Language Models: Principles, Taxonomy, Challenges, and Open Questions},
  journal   = {ACM Transactions on Information Systems},
  volume    = {43},
  number    = {2},
  pages     = {1--55},
  year      = {2025},
  doi       = {10.1145/3703155}
}

@inproceedings{ji2023mitigating,
  author    = {Ji, Ziwei and Yu, Tiezheng and Xu, Yan and Lee, Nayeon and Ishii, Etsuko and Fung, Pascale},
  title     = {Towards Mitigating Hallucination in Large Language Models via Self-Reflection},
  booktitle = {Findings of the Association for Computational Linguistics: EMNLP 2023},
  pages     = {1827--1843},
  year      = {2023},
  publisher = {Association for Computational Linguistics},
  address   = {Singapore},
  doi       = {10.18653/v1/2023.findings-emnlp.123},
  url       = {https://aclanthology.org/2023.findings-emnlp.123}
}

@article{pardos2024chatgpt,
  author    = {Pardos, Zachary A. and Bhandari, Shreya},
  title     = {ChatGPT-generated help produces learning gains equivalent to human tutor-authored help on mathematics skills},
  journal   = {PLOS ONE},
  volume    = {19},
  number    = {5},
  pages     = {e0304013},
  year      = {2024},
  doi       = {10.1371/journal.pone.0304013},
  publisher = {Public Library of Science}
}

@inproceedings{li2024dawn,
  author    = {Li, Junyi and Chen, Jie and Ren, Ruiyang and Cheng, Xiaoxue and Zhao, Xin and Nie, Jian-Yun and Wen, Ji-Rong},
  title     = {The Dawn After the Dark: An Empirical Study on Factuality Hallucination in Large Language Models},
  booktitle = {Proceedings of the 62nd Annual Meeting of the Association for Computational Linguistics (Volume 1: Long Papers)},
  pages     = {10879--10899},
  year      = {2024},
  address   = {Bangkok, Thailand},
  publisher = {Association for Computational Linguistics},
  doi       = {10.18653/v1/2024.acl-long.586},
  url       = {https://aclanthology.org/2024.acl-long.586}
}

@techreport{apa2015top20,
  author       = {{American Psychological Association, Coalition for Psychology in Schools and Education}},
  title        = {Top 20 Principles from Psychology for PreK--12 Teaching and Learning},
  institution  = {American Psychological Association},
  type         = {Technical Report},
  year         = {2015},
  url          = {https://www.apa.org/ed/schools/cpse/top-twenty-principles.pdf},
}

@article{abuhassna2024exploring,
  title={Exploring the synergy between instructional design models and learning theories: A systematic literature review},
  author={Abuhassna, Hassan and Adnan, Mohamad Azrien Bin Mohamed and Awae, Fareed},
  journal={Contemporary Educational Technology},
  volume={16},
  number={2},
  pages={ep499},
  year={2024},
  publisher={Bastas}
}

@article{Scott2020DesignBasedResearch,
  author       = {Scott, Emily E. and Wenderoth, Mary Pat and Doherty, Jennifer H.},
  title        = {Design‐Based Research: A Methodology to Extend and Enrich Biology Education Research},
  journal      = {CBE—Life Sciences Education},
  volume       = {19},
  number       = {3},
  pages        = {es11},
  year         = {2020},
  doi          = {10.1187/cbe.19-11-0245},
  url          = {https://doi.org/10.1187/cbe.19-11-0245}
}

@book{bakker2004design,
  title     = {Design Research in Statistics Education: On Symbolizing and Computer Tools},
  author    = {Bakker, Arthur},
  year      = {2004},
  publisher = {CD-B\`eta Press},
  address   = {Utrecht, The Netherlands}
}

@article{lee2025collaborative,
  title={Collaborative learning with artificial intelligence speakers: pre-service elementary science teachers’ responses to the prototype},
  author={Lee, Gyeong-Geon and Mun, Seonyeong and Shin, Myeong-Kyeong and Zhai, Xiaoming},
  journal={Science \& Education},
  volume={34},
  number={2},
  pages={847--875},
  year={2025},
  publisher={Springer}
}

@article{cobb2012analyzing,
  title={Analyzing educational policies: A learning design perspective},
  author={Cobb, Paul and Jackson, Kara},
  journal={Journal of the Learning Sciences},
  volume={21},
  number={4},
  pages={487--521},
  year={2012},
  publisher={Taylor \& Francis}
}

@article{celik2025co,
  title={Co-constructing adaptive lesson plans with GenAI: Pre-service teachers' Intelligent-TPACK and prompt engineering strategies},
  author={Celik, Ismail and Kontkanen, Sini and Laru, Jari and Dalyanci, Alanur Ahsen},
  journal={Computers \& Education},
  pages={105485},
  year={2025},
  publisher={Elsevier}
}

@article{bhandari2024evaluating,
  title={Evaluating the psychometric properties of ChatGPT-generated questions},
  author={Bhandari, Shreya and Liu, Yunting and Kwak, Yerin and Pardos, Zachary A},
  journal={Computers and Education: Artificial Intelligence},
  volume={7},
  pages={100284},
  year={2024},
  publisher={Elsevier}
}

@inproceedings{jin2024teach,
  title={Teach ai how to code: Using large language models as teachable agents for programming education},
  author={Jin, Hyoungwook and Lee, Seonghee and Shin, Hyungyu and Kim, Juho},
  booktitle={Proceedings of the 2024 CHI Conference on Human Factors in Computing Systems},
  pages={1--28},
  year={2024}
}

@inproceedings{lu2024generative,
  title={Generative students: Using llm-simulated student profiles to support question item evaluation},
  author={Lu, Xinyi and Wang, Xu},
  booktitle={Proceedings of the Eleventh ACM Conference on Learning@ Scale},
  pages={16--27},
  year={2024}
}

@article{li2024evolving,
  title={Evolving agents: Interactive simulation of dynamic and diverse human personalities},
  author={Li, Jiale and Li, Jiayang and Chen, Jiahao and Li, Yifan and Wang, Shijie and Zhou, Hugo and Ye, Minjun and Su, Yunsheng},
  journal={arXiv preprint arXiv:2404.02718},
  year={2024}
}

@inproceedings{jin2025teachtune,
  title={Teachtune: Reviewing pedagogical agents against diverse student profiles with simulated students},
  author={Jin, Hyoungwook and Yoo, Minju and Park, Jeongeon and Lee, Yokyung and Wang, Xu and Kim, Juho},
  booktitle={Proceedings of the 2025 CHI Conference on Human Factors in Computing Systems},
  pages={1--28},
  year={2025}
}

@inproceedings{nguyen2024simulating,
  title={Simulating climate change discussion with large language models: Considerations for science communication at scale},
  author={Nguyen, Ha and Nguyen, Victoria and L{\'o}pez-Fierro, Sar{\'\i}ah and Ludovise, Sara and Santagata, Rossella},
  booktitle={Proceedings of the eleventh ACM conference on learning@ scale},
  pages={28--38},
  year={2024}
}

@article{goslen2025llm,
  title={Llm-based student plan generation for adaptive scaffolding in game-based learning environments},
  author={Goslen, Alex and Kim, Yeo Jin and Rowe, Jonathan and Lester, James},
  journal={International journal of artificial intelligence in education},
  volume={35},
  number={2},
  pages={533--558},
  year={2025},
  publisher={Springer}
}

@incollection{hoadley2018short,
  title={A short history of the learning sciences},
  author={Hoadley, Christopher},
  booktitle={International handbook of the learning sciences},
  pages={11--23},
  year={2018},
  publisher={Routledge}
}

@article{saccak2022down,
  title={Down the rabbit hole: Revisiting etymology, epistemology, history and practice of instructional and learning design},
  author={Sa{\c{c}}ak, Beg{\"u}m and Bozkurt, Aras and Wagner, Ellen},
  journal={ELearn},
  volume={2022},
  number={3},
  year={2022},
  publisher={ACM New York, NY, USA}
}

@book{smith2004instructional,
  title={Instructional design},
  author={Smith, Patricia L and Ragan, Tillman J},
  year={2004},
  publisher={John Wiley \& Sons}
}

@article{heaster2020popular,
  title={Popular Instructional Design Models: Their Theoretical Roots and Cultural Considerations.},
  author={Heaster-Ekholm, Kristen Lina},
  journal={International Journal of Education and Development using Information and Communication Technology},
  volume={16},
  number={3},
  pages={50--65},
  year={2020},
  publisher={ERIC}
}

@article{nathan2010learning,
  title={Learning sciences},
  author={Nathan, Mitchell J and Wagner Alibali, Martha},
  journal={Wiley Interdisciplinary Reviews: Cognitive Science},
  volume={1},
  number={3},
  pages={329--345},
  year={2010},
  publisher={Wiley Online Library}
}

@article{ishimaru2022transforming,
  title={Transforming the role of RPPs in remaking educational systems},
  author={Ishimaru, Ann M and Barajas-L{\'o}pez, Filiberto and Sun, Min and Scarlett, Keisha and Anderson, Eric},
  journal={Educational Researcher},
  volume={51},
  number={7},
  pages={465--473},
  year={2022},
  publisher={SAGE Publications Sage CA: Los Angeles, CA}
}

@article{brown1992design,
  title={Design experiments: Theoretical and methodological challenges in creating complex interventions in classroom settings},
  author={Brown, Ann L},
  journal={The journal of the learning sciences},
  volume={2},
  number={2},
  pages={141--178},
  year={1992},
  publisher={Taylor \& Francis}
}

@book{piaget1970,
  title={Science of education and the psychology of the child},
  author={Piaget, Jean},
  year={1970},
  publisher={Viking Press},
  address={New York}
}

@book{vygotsky1978,
  title={Mind in society: The development of higher psychological processes},
  author={Vygotsky, Lev S.},
  year={1978},
  publisher={Harvard University Press},
  address={Cambridge, MA}
}

@article{sawyer2022introduction,
  title={An introduction to the learning sciences},
  author={Sawyer, R Keith},
  journal={The Cambridge handbook of the learning sciences},
  volume={3},
  pages={1--24},
  year={2022},
  publisher={Cambridge University Press}
}

@article{yoon2017learning,
  title={What do learning scientists do? A survey of the ISLS membership},
  author={Yoon, Susan A and Hmelo-Silver, Cindy E},
  journal={Journal of the Learning Sciences},
  volume={26},
  number={2},
  pages={167--183},
  year={2017},
  publisher={Taylor \& Francis}
}

@article{gronseth2024bridging,
  title={Bridging Silos: Collaborating to Create Authentic Learning Experiences for Future Instructional Designers},
  author={Gronseth, Susie L and Itani, Amani and Beech, Bettina M and Heitman, Elizabeth and Bruce, Marino A and Kakadiaris, Ioannis A},
  journal={The Journal of Applied Instructional Design},
  volume={13},
  number={1},
  pages={21--33},
  year={2024}
}

@article{iqbal2025codesigning,
  title={Co-Designing an {AI} Literacy Curriculum for Elementary Education Using Design Thinking},
  author={Iqbal, R. A. and Islam, R. S. and Cockerham, D.},
  journal={The Journal of Applied Instructional Design},
  year={2025},
  volume={August 2025},
  doi={10.59668/2223.22436},
  url={https://doi.org/10.59668/2223.22436}
}

@article{ritzhaupt2015knowledge,
  title={Knowledge and skills needed by instructional designers in higher education},
  author={Ritzhaupt, Albert D and Kumar, Swapna},
  journal={Performance Improvement Quarterly},
  volume={28},
  number={3},
  pages={51--69},
  year={2015},
  publisher={Wiley Online Library}
}

@article{kenny2005review,
  title={A review of what instructional designers do: Questions answered and questions not asked},
  author={Kenny, Richard and Zhang, Zuochen and Schwier, Richard and Campbell, Katy},
  journal={Canadian Journal of Learning and Technology/La revue canadienne de l’apprentissage et de la technologie},
  volume={31},
  number={1},
  year={2005},
  publisher={Canadian Network for Innovation in Education}
}

@article{miller2016finding,
  title={Finding our voice: Instructional designers in higher education},
  author={Miller, Sandie and Stein, Gayle},
  journal={EDUCAUSE Review (Online)},
  year={2016},
  publisher={Educause}
}

@article{pan2009exploring,
  title={Exploring dynamics between instructional designers and higher education faculty: An ethnographic case study},
  author={Pan, Cheng-Chang Sam and Thompson, Kelvin},
  journal={Journal of Educational Technology Development and Exchange (JETDE)},
  volume={2},
  number={1},
  pages={3},
  year={2009}
}

@article{park2017refining,
  title={Refining a competency model for instructional designers in the context of online higher education.},
  author={Park, Jae-Young and Luo, Heng},
  journal={International Education Studies},
  volume={10},
  number={9},
  pages={87--98},
  year={2017},
  publisher={ERIC}
}

@inproceedings{you2010instructional,
  title={Instructional designers’ role in assisting instructors in the implementation of best practices in distance learning course design and delivery in higher education: Instructors’ perspectives},
  author={You, Peter and Teclehaimanot, Berhane},
  booktitle={eLearn: World Conference on EdTech},
  pages={857--865},
  year={2010},
  organization={Association for the Advancement of Computing in Education (AACE)}
}

@article{scoppio2017mind,
  title={Mind the gap: Enabling online faculty and instructional designers in mapping new models for quality online courses},
  author={Scoppio, Grazia and Luyt, Ilka},
  journal={Education and Information Technologies},
  volume={22},
  number={3},
  pages={725--746},
  year={2017},
  publisher={Springer}
}

@article{olesova2019impact,
  title={The Impact of the Cooperative Mentorship Model on Faculty Preparedness to Develop Online Courses.},
  author={Olesova, Larisa and Campbell, Susan},
  journal={Online Learning},
  volume={23},
  number={4},
  pages={192--213},
  year={2019},
  publisher={ERIC}
}

@article{muljana2021utilizing,
  title={Utilizing learning analytics in course design: Voices from instructional designers in higher education},
  author={Muljana, Pauline Salim and Luo, Tian},
  journal={Journal of Computing in Higher Education},
  volume={33},
  number={1},
  pages={206--234},
  year={2021},
  publisher={Springer}
}

@article{chen2024framework,
  title={A framework for infrastructuring sustainable innovations in education},
  author={Chen, Bodong},
  journal={Journal of the Learning Sciences},
  volume={33},
  number={3},
  pages={583--612},
  year={2024},
  publisher={Taylor \& Francis}
}

@incollection{yanchar2009beyond,
  title={Beyond the theory-practice split in instructional design: The current situation and future directions},
  author={Yanchar, Stephen C and South, Joseph B},
  booktitle={Educational media and technology yearbook},
  pages={81--100},
  year={2009},
  publisher={Springer}
}

@inproceedings{zhou2022large,
  title={Large language models are human-level prompt engineers},
  author={Zhou, Yongchao and Muresanu, Andrei Ioan and Han, Ziwen and Paster, Keiran and Pitis, Silviu and Chan, Harris and Ba, Jimmy},
  booktitle={The eleventh international conference on learning representations},
  year={2022}
}

@article{lewis2020retrieval,
  title={Retrieval-augmented generation for knowledge-intensive nlp tasks},
  author={Lewis, Patrick and Perez, Ethan and Piktus, Aleksandra and Petroni, Fabio and Karpukhin, Vladimir and Goyal, Naman and K{\"u}ttler, Heinrich and Lewis, Mike and Yih, Wen-tau and Rockt{\"a}schel, Tim and others},
  journal={Advances in neural information processing systems},
  volume={33},
  pages={9459--9474},
  year={2020}
}

@article{mcintyre2005bridging,
  title={Bridging the gap between research and practice},
  author={McIntyre, Donald},
  journal={Cambridge journal of education},
  volume={35},
  number={3},
  pages={357--382},
  year={2005},
  publisher={Taylor \& Francis}
}

@misc{schlicht2024bridging,
  title={Bridging the research-practice gap in education: Initiatives from three OECD countries},
  author={Schlicht-Schm{\"a}lzle, R and Hill, J and Folkvord, KA and Tharaldsen, KB and Wargo, J and Hartmann, U and R{\'e}vai, N},
  year={2024},
  publisher={OECD Directorate for Education and Skills Working Paper}
}

@incollection{doyle2013ecological,
  title={Ecological approaches to classroom management},
  author={Doyle, Walter},
  booktitle={Handbook of classroom management},
  pages={107--136},
  year={2013},
  publisher={Routledge}
}

@article{ferrara2023should,
  title={Should chatgpt be biased? challenges and risks of bias in large language models},
  author={Ferrara, Emilio},
  journal={arXiv preprint arXiv:2304.03738},
  year={2023}
}

@article{mah2025co,
  title={A co-creative approach for AI-enhanced instructional design: Combining generative artificial intelligence and learning analytics},
  author={Mah, Dana-Kristin and Egloffstein, Marc},
  journal={The Journal of Applied Instructional Design: JAID},
  volume={14},
  number={3},
  pages={1--19},
  year={2025},
  publisher={EdTech Books}
}

@article{wang2005design,
  title={Design-based research and technology-enhanced learning environments},
  author={Wang, Feng and Hannafin, Michael J},
  journal={Educational technology research and development},
  volume={53},
  number={4},
  pages={5--23},
  year={2005},
  publisher={Springer}
}

@article{edelson2002design,
  title={Design research: What we learn when we engage in design},
  author={Edelson, Daniel C},
  journal={The Journal of the Learning sciences},
  volume={11},
  number={1},
  pages={105--121},
  year={2002},
  publisher={Taylor \& Francis}
}

@article{hu2025llm,
  title        = {Exploring the potential of LLM to enhance teaching plans through teaching simulation},
  author       = {Hu, Bihao and Zhu, Jiayi and Pei, Yiying and Gu, Xiaoqing},
  journal      = {npj Science of Learning},
  volume       = {10},
  number       = {1},
  pages        = {7},
  year         = {2025},
  publisher    = {Nature Publishing Group},
  doi          = {10.1038/s41539-025-00300-x}
}

@article{yan2021including,
  title={Including learning analytics in the loop of self-paced online course learning design},
  author={Yan, Hongxin and Lin, Fuhua and Kinshuk},
  journal={International Journal of Artificial Intelligence in Education},
  volume={31},
  number={4},
  pages={878--895},
  year={2021},
  publisher={Springer}
}

@article{celik2026knowledge,
  title={Knowledge construction with GenAI: The role of theory-informed prompt engineering in achieving pedagogical alignment},
  author={Celik, Ismail and Zabolotna, Kateryna and Viberg, Olga},
  journal={Innovations in Education and Teaching International},
  pages={1--19},
  year={2026},
  publisher={Taylor \& Francis}
}

@article{coburn2013practice,
  title={Practice partnerships: A strategy for leveraging research for educational improvement in school districts.},
  author={Coburn, Cynthia E and Penuel, William R and Geil, Kimberly E},
  journal={William T. Grant Foundation},
  year={2013},
  publisher={ERIC}
}

@article{coburn2016research,
  title={Research--practice partnerships in education: Outcomes, dynamics, and open questions},
  author={Coburn, Cynthia E and Penuel, William R},
  journal={Educational researcher},
  volume={45},
  number={1},
  pages={48--54},
  year={2016},
  publisher={Sage Publications Sage CA: Los Angeles, CA}
}

@book{gasevic2016we,
  title={How do we start? State and directions of learning analyics adoption},
  author={Gasevic, Dragan and Dawson, Shane and Pardo, Abelardo},
  year={2016},
  publisher={International Council for Open and Distance Education}
}

@article{wong2020review,
  title={A review of learning analytics intervention in higher education (2011--2018)},
  author={Wong, Billy Tak-ming and Li, Kam Cheong},
  journal={Journal of Computers in Education},
  volume={7},
  number={1},
  pages={7--28},
  year={2020},
  publisher={Springer}
}

@article{raj2024improved,
  title={An improved adaptive learning path recommendation model driven by real-time learning analytics},
  author={Raj, Nisha S and Renumol, VG},
  journal={Journal of Computers in Education},
  volume={11},
  number={1},
  pages={121--148},
  year={2024},
  publisher={Springer}
}

@inproceedings{li2025turning,
  title={Turning real-time analytics into adaptive scaffolds for self-regulated learning using generative artificial intelligence},
  author={Li, Tongguang and Nath, Debarshi and Cheng, Yixin and Fan, Yizhou and Li, Xinyu and Rakovi{\'c}, Mladen and Khosravi, Hassan and Swiecki, Zachari and Tsai, Yi-Shan and Ga{\v{s}}evi{\'c}, Dragan},
  booktitle={Proceedings of the 15th international learning analytics and knowledge conference},
  pages={667--679},
  year={2025}
}

@article{liu2025effectiveness,
  title={The effectiveness of learning analytics-based interventions in enhancing students’ learning effect: A meta-analysis of empirical studies},
  author={Liu, Yan and Wang, Wei and Xu, Enwei},
  journal={SAGE Open},
  volume={15},
  number={2},
  pages={21582440251336707},
  year={2025},
  publisher={SAGE Publications Sage CA: Los Angeles, CA}
}

@article{fishman2004creating,
  title={Creating a Framework for Research on Systemic Technology Innovations},
  author={Fishman, Barry and Marx, Ronald W and Blumenfeld, Phyllis and Krajcik, Joseph and Soloway, Elliot},
  journal={The Journal of the Learning Sciences},
  volume={13},
  number={1},
  pages={43--76},
  year={2004},
  publisher={Taylor \& Francis}
}

\end{document}